# An Optimal Prefix Replication Strategy for VoD Services

M Dakshayini[1], Dr T R GopalaKrishnan Nair[2].

**Abstract**— In this paper we propose scalable proxy servers cluster architecture of interconnected proxy servers for high quality and high availability services. We also propose an optimal regional popularity based video prefix replication strategy and a scene change based replica caching algorithm that utilizes the zipf-like video popularity distribution to maximize the availability of videos closer to the client and request-servicing rate thereby reducing the client rejection ratio and the response time for the client. The simulation results of our proposed architecture and algorithm show the greater achievement in maximizing the availability of videos, client request-servicing rate and in reduction of initial start-up latency and client rejection ratio.

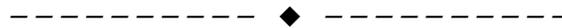

## 1 INTRODUCTION

With the advanced technologies developed in the areas of multimedia and high speed networks, video-on-demand (VoD) service is considered as the promising trend in home entertainment, as well as in banking, education, home shopping, and interactive games[4]. The vivid growth of World Wide Web has increased the use of communication networks in accessing www and also the user latency, network congestion. Owing to these problems, deploying a proxy server with the prediction of satisfying future client requests without communicating with the main server directly can reduce the response time delay, server load, and network traffic substantially. Proxy server is closely located to clients. When a client wants some data, proxy server intercepts the request and provides the data if possible. Otherwise it contacts the remote main server, and then saves the received data in its cache and forwards them to the client. As a result, the response time is significantly reduced, and the network utilization can be improved since the communication with the main server decreases. Further by interconnecting the group of proxy servers and sharing the data among these inter connected proxy servers can still improve the client acceptance rate and reduce the response time, network traffic and client rejection rate. Video data has several different characteristics when compared with other media such as text, image, and audio, etc. First, video data requires a huge amount of storage compared with text, image, and audio.

Thus it is not efficient to store the whole data of video object at proxy server. Second, the generic attribute of video is highly variable bit rate (VBR). Accordingly the required client's buffer size is very large. Third, it is time-constrained. Actually, video is a principal element in the performance of proxy server. The compressed video data is generally variable-bit-rate due to the generic characteristics of entropy coder, frame type and scene change motion change of the underlying video. Assuming apriori knowledge about the video popularities, we propose an efficient video placement method with prefix replication.

The major problems are how to determine the average number of replicas (replication degree) of videos, size of the replica and the way of placing the video replicas on the distributed storage cluster for high quality, high availability and load sharing under resource constraints. High availability in this context has two meanings: low rejection rate, high replication degree and client acceptance rate. The objective of load sharing is to improve system throughput in rush-hours and hence reduce the rejection rate and increase the client acceptance rate. In this paper we present a novel load sharing architecture and an optimal video prefix replication and prefix (replica) placement algorithm. Specifically,

1. We formulate the video prefix replication and placement among the PSs of LPSG for high service quality and high service availability as a combinatorial optimization problem.
2. We propose an optimal video prefix replication

- M.Dakshayini is Research Scholar, Dr. MGR University,Chenni*. Working with the Dept. of Information Scienc and Engg, BMSCE, Bangalore, India.*
- *Dr.T.R.Gopalakrishnan Nair is the Director of Research and Industry Incubation Centre, DSI,Bangalore,India*

algorithm and a bounded prefix caching algorithm for videos based on their regional popularity.

3. We propose an efficient prefix caching algorithm. We conduct an ample performance evaluation of the algorithms and demonstrate their efficiency via simulations.

The rest of the paper is organized as follows: Section 2 presents the related work. In section 3 we present a model and formulation of the proposed problem, Section 4 describes the proposed approach and algorithms in detail, In section 5 we present a simulation model, Section 6 presents the simulation results and comparison of our proposed regional popularity based replication and placement algorithm (RPR-P) with zifpf replication with smallest load first (zipfR-SLFA) and classical replication with round robin placement (CR-RR) algorithms, Finally, Finally in section 7 we conclude the paper.

## 2 RELATED WORK

This section briefly discusses the previous work as follows, several works on VoD applications focused on scalability and reliability of distributed VoD servers [5], [9], [10]. The video replication and placement problem has been studied extensively in many papers [3], [5], [6]. Tay and pang [6] have proposed an algorithm in called *GWQ* (Global waiting queue) which reduces the initial startup delay by sharing the videos in a distributed loosely coupled *VoD* system by balancing the load between the lightly loaded proxy servers and heavily loaded proxy servers in a distributed *VoD*. In [6] Sonia Gonzalez, Navarro, Zapata proposed a more realistic partial replication and load sharing algorithm PRLS to distribute the load in a distributed *VoD* system. Xiabo Zhou, Cheng-Zhong xu in [2] have proposed an algorithm (zipfR-SLFA) for video replication and placement algorithm, which utilizes the information about Zipf-like video popularity distribution. They have replicated the complete videos evenly in all the servers, for which the storage capacity of individual proxy server should be very large to store all the videos. This may not allow each server to store replicas of more number of videos. Our proposed scheme replicates regional (local and global) popularity based prefix of the videos using scene change based buffer allocation algorithm, there by utilizing the proxy server storage space more efficiently to store replicas of more number of videos. In this paper, we present an architecture of distributed proxy servers, a Tracker[TR] and clients for serving videos with a target to optimize the Request-Service delay time. This architecture consists of a Main multimedia server [*MMS*] which is far away from the user and is connected to a set of trackers. Each tracker is in turn connected to a group of proxy servers and these proxy servers are assumed to be interconnected in a ring pattern, this arrangement of cluster of proxy servers is called as Local Proxy servers Group[*LPSG|L_p*]. Each of such *LPSG* is connected to

*MMS* and is in turn connected to its left and right neighboring *LPSG* in a ring fashion through its tracker. We also propose an efficient scene change based cache allocation algorithm at the proxy server to allocate the cache according to the size of P and B frames and popularity of the video. The main goal of this algorithm is to allow LPSG system to store prefixes of more number of videos.

## 3 THE OPTIMIZATION PROBLEM

### 3.1 The Model

We consider a cluster of $M$ proxy servers, $PS=(PS1,PS2,…PS_m)$, and a set of $N$ various videos, V = {$V1,V2,…V_n$}. Every $i^{th}$ video $V_i$ is divided in 3 parts, first $W_1$ minutes of each video $V_i$ is referred to as *prefix -1* of $V_i$ (*pref-1*)$_i$ and is considered as replica of $V_i$ to be replicated at L selected PSs. Next $W_2$ minutes of video $V_i$ is referred to as *prefix-2* of $V_i$ (*pref-2*)$_i$ and is cached at TR of LPSG. The rest of the video is referred to as suffix of the video and the complete video is stored at *MMS*. We consider that, the size(duration) of the complete video (S) $V_i$ (i=1..n) in the set $V$ can range from 30 minutes to 150 minutes. The sizes (duration) of (*pref-1*)$_i$ and (*pref-2*)$_i$ as $d_1$and $d_2$ respectively. These sizes vary according to the local popularity of the video, say $d_1$ and $d_2$ can range from 20 minutes to 120 minutes based on their local popularity. Let the encoding rate of the video $v_i$ is $b_i$hence the caching space for video $v_i$is calculated as $c_i = d_i.b_i$. Popularity of the videos varies with the number of hits to the videos. We consider the replication of the video *pref-1* and SC-caching for the peak period of length $d_1$. This is because one of the main objectives of our prefix replication and caching is to provide high service availability during the peak period. Load sharing is critical for improving the throughput and service availability during the peak period. We make the following assumptions, regarding the video relative popularity distributions and the request arrival rates.

1. It is assumed that, the popularity of the videos $p_i$ is known before the prefix replication and caching. The relative popularity of the videos follows Zipf-like distributions with a skew parameter of ѳ typically, $0.271 \le \theta \le 1$ [1]. The probability of choosing the ith video is

   $$p_i = i^{-\theta} / \sum_{j=1}^{N} j^{-\theta}$$

2. We assume that client's requests (X/hr) arrive according to Poisson process with $\lambda$ as shown in the simulation model Figure 2. Let $S_i$ be the size (duration in minutes) of $i^{th}$ video(i=1..N) with mean arrival rates $\lambda_1 . . . \lambda_N$ respectively that are being streamed to the users using $M$ proxy servers ($PSs$) of Z $LPSGs$ ($L_p$ p=1..Z).

### 3.2 Formulation of the Problem

The main objective of the video prefix replication and SC-caching algorithm at the PS, which behaves like an intermediate node between the client and the MMS, is to have high service quality and high service availability by utilizing the buffer and bandwidth among the PSs of LPSG. Replication of the prefix among the PSs of LPSG enhances the availability of videos and request-servicing ability of the proposed system using SC-caching. Load sharing mechanism improves the system throughput and hence the service availability. Increasing the prefix size achieves high quality but decreases the replication degree due to the storage constraint. In this scenario, MMS-to-proxy channel provides guaranteed constant bandwidth service and this higher bandwidth service is provided at the higher cost. On the other hand, PS-to-client channel is fast and reliable at the even lower and fixed price. Thus the initial access delay ($I^a d$) and the cost between proxy server and client are negligeble compared with those between MMS and PS, and the cached data is almost instantly available at the proxy server to serve the user's request immediately as it arrives. Thus, the required client's buffer size on the PS-to-client channel is almost negligeble compared with that in the MMS-to-client channel. The objective of prefix replication and SC-caching hence is to maximize the prefix size based on the local popularity of the video and the average number of replicas (replication degree), and to minimize the client rejection rate in the system. Depending on the probability of occurrence of user requests to any video, the popularity of the videos and size of prefix to be cached at *PS* and *TR is* determined.

i.e. $d_1$ and $d_2$ $\alpha$ $n_i$

So $\quad d_1 = x_i \times s_i$ where $0 < x_i < 1$

$\quad\quad d_2 = x_i \times (s_i \cdot d_1)$ where $0 < x_i < 1$

Where
$d_1$ − Size of the prefix-1
$d_2$ − Size of the prefix-2
$n_i$ - is the total number of requests for $i^{th}$ video $v_i$.
$x_i$ - The probability of requests arrival for the $i^{th}$ video from last t minutes.
$s_i$ - The size(duration) of the complete video

And another output stochastic variable $I^a d$ is the average initial access delay for all the requested videos at LPSG. Thus $I^a d$ is a sample mean of response times $I^a d_1$, $I^a d_2$..... $I^a d_m$ at all *PSs* $PS_1$, $PS_2$, $PS_m$.

The main objective of our proposal is to maximize the system throughput by maximizing the service availability and minimizing the request rejection ratio $R_{rej}$, average initial access delay $I^a d$ and average cache miss rate. This can be formulated as follows.

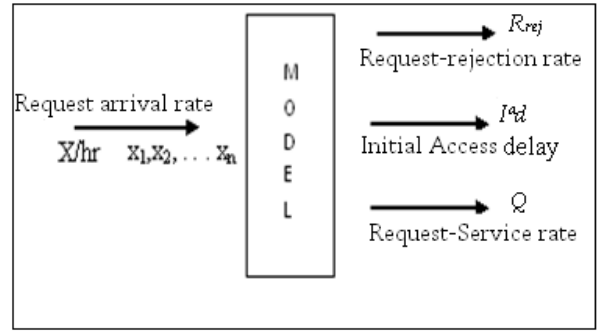

Fig. 1. System Model

Max Obj $\quad Q = \dfrac{1}{M} \sum_{i=1}^{M} Q_i$ $\hspace{2cm}$ (1)

Min Obj $\quad I^a d = \dfrac{1}{M} \sum_{j=1}^{M} \left( \dfrac{1}{Q} \sum_{i=1}^{Q} (I^a d)_i \right)$ , and

$\quad\quad R_{rej} = \left( \dfrac{1}{M} \sum_{i=1}^{M} (N_{rej})_i \right) \Big/ (R/M)$ $\hspace{1cm}$ (2)

Where $Q$ is the total number of requests served at $L_P$.
$\quad$ $R$ is the total number of requests arrived at $L_P$.
$\quad$ $N_{rej}$ is the number of requests rejected at a specific PS.

These objectives are subject to the following constraints: (1) PS and TR storage capacity (2) distribution of all the replicas of an individual video to different servers. Let $v_i^j$ be the index of the server (j) on which replica of $i^{th}$ video is placed. Specifically, we give storage constraints from the perspective of proxy server and tracker

$B = \sum_{i=1}^{K} (pref - 1)_i$ , $\quad P = \sum_{i=1}^{H} (pref - 2)_i$ $\quad d_1, d_2 > 0$ $\quad$ (3)

The tracker and each proxy server in the cluster has the caching capacity of large enough to cache total $P$ and $B$ minutes of $H$ and $K$ number of videos respectively

The second constraint is the requirement of distribution of all replicas of an individual video to different proxy servers. That is, all ri replicas of video vi must be placed on ri PSs. Specifically,

$v_i^{j1} \neq v_i^{j2}$ $\quad 1 \leq j_1, j_2 \leq r_i,$ $\quad j1 \neq j2.$ $\hspace{1cm}$ (4)

Note that if multiple replicas of a video are placed to the same server, it implies that duplicate copy must be deleted. For this reason, we have one more replication constraint

$1 \leq ri \leq M$ $\quad$ for every $v_i \in V$ $\hspace{1cm}$ (5)

In summary, we formulate the video prefix replication and placement problem as a maximization of Eq(1) and minimization of Eq(2) subject to constraints Eq(3) to Eq(5).

# 4 THE OPTIMIZATION PROBLEM

## 4.1 The Prefix replication and Placement

The proposed VoD architecture for our problem is as shown in Fig.2. This architecture consists of a *MMS*, which is connected to a group of trackers (*TRs*), Each *TR* has various modules like

*Communication Module (CM)* – it communicates with the *PS* and *MMS*.

*Service Manager (SM$_{TR}$)* – this handles and manages the requests from the *PS*.

*Database* – stores the complete details of presence and size of replica *(pref-1)* of videos at all the *PSs*.

*Video distributing Manager (VDM)* – is responsible for deciding the video prefixes, and sizes of the replica *(pref-1) and (pref-2)* of videos to be cached at PS and TR respectively. Also handles the distribution and management of these videos among the *PSs* of LPSG, based on video's global and local popularity.

Each *TR* is in turn connected to a set of *PSs*. These *PSs* are connected among themselves in a ring fashion.

Each *PS* has various modules such as,

*Interaction Module (IM)* – Interacts with the user and *TR*.

*Service Manager (SM$_{PS}$)* – Handles the requests from the user,

*Popularity agent (PA)* – Observes and updates the popularity of videos at PS as well as at *TR,*

*Cache Allocator (CA)* – Allocates the cache blocks to various frames of the video prefix.

Also to each of these PSs a large number of users are connected [*LPSG*]. Each PS is called as a parent PS to its clients. All these *LPSGs* are interconnected through their *TR* in a ring pattern. The *PS* caches the *(pref-1)* of videos distributed by *VDM*, and streams this cached portion of video to the client upon the request through *LAN* using its less expensive bandwidth.

Replication and placement of the *(pref-1)* and *(pref-2)* of the videos is done as follows. If *V$_i$* is popular at all PSs of LPSG (globally popular) then TR calculates the size *d$_1$* of the replica *pref-1*, *d$_2$* of *pref-2* and *pref-1* is replicated at all the M PSs.

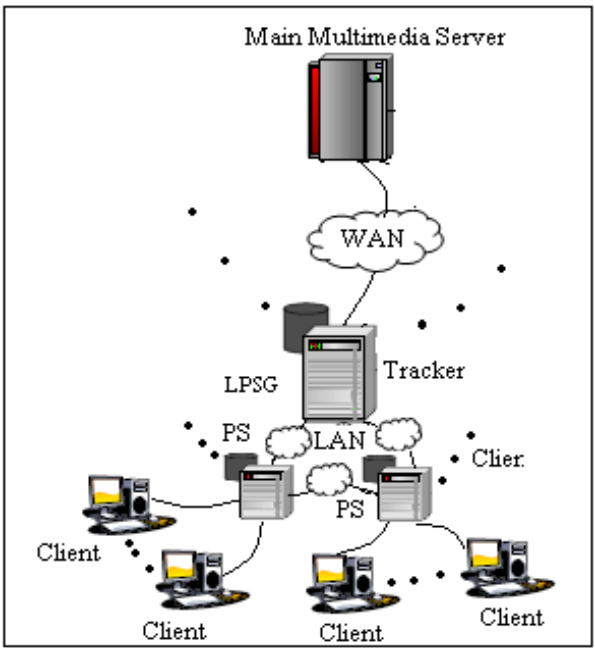

Fig. 2. Proposed VoD Architecture

Otherwise it is replicated only across selected *L PSs* of LPSG, in which the frequency of accessing the video *V$_i$* is very high and has enough storage space to accommodate *(pref-1)*. This arrangement of replicating the popularity based *(pref-1)* at *L PSs* increases the availability of videos and hence helps the system to provide the serve immediately as the request arrives. This in turn increases the client service rate.

### Algorithm : Regional popularity based *prefix-1* replication and Placement

[Nomenclature

$N(r_i)$ : Number of replicas of i$^{th}$ video

$x_{vi}^{k}$ : probability of requests arrival for video $v_i$ at PS k

$d_1^{vi}$ : Size of the replica of i$^{th}$ video

$t_P$ : Threshold popularity.

$B_{avl}^{z}$ : Buffer available at the *PS z*]

For i = 1 to N

{

  **if** ($v_i$ is globally popular)

   {

    $N(r_i) = M$

    *VDM* calculates the sizes $d_1^{vi}$, $d_2^{vi}$.

$$d_1^{vi} = \frac{1}{M}\sum_{k=1}^{M} x_{vi}^{k} \times s_i \quad \text{where} \quad 0 < x_i < 1$$

$$d_2^{vi} = \frac{1}{M}\sum_{k=1}^{M} x_{vi}^{k} \times (s_i - d_1^{vi}) \quad \text{where} \quad 0 < x_i < 1$$

    $v_i = (pref-1)_i + (pref-2)_i + (suffi\ x)_i$

    Replicates the *(pref-1)$_{vi}$* at all *M PSs* using *SC-caching*

  }

  **else** if($v_i$ is popular only at some regions)

   L=1;

   **for** z = 1 to M

   {

    **if** ($pop(v_i)^z > t_P$)

    **if** ( $B_{avl}^{z} \geq d_1^{vi}$ )

    {

     L++;

     $d_1^{vi} = x_{vi}^{z} \times s_i \quad \text{where} \quad 0 < x_i < 1$

    }

    Replicate *(pref-1)$_{vi}$* at selected *L PSs* using *SC-caching*

    **else**

     share the *(pref-1)$_{vi}$* stored at other PSs of LPSG

    $d_2^{vi} = max( x_{vi}^{1}, x_{vi}^{2}, ..., x_{vi}^{M}) \times (s_i - d_1^{vi}) \quad \text{where} \quad 0 < x_i < 1$

    Cache the (pref-2)$_{vi}$ at TR using *SC-caching*

  }

}

## 4.2 SC- Prefix Caching Algorithm

The SC-prefix caching algorithm is proposed to maximize the storage capacity of the system by allocating the cache blocks dynamically based on the current size of I, B and P frames. This algorithm utilizes the VBR attribute of the videos. When the difference between the cache requirements of two consecutive frames is very large, the process of buffer reallocation is initiated. This scheme improves buffer utilization significantly compared to the concept of static buffer allocation for the entire scene. A scene change is identified if the change in the number of bits between successive frames exceeds a certain threshold in a continuous manner. Denote $F_n$ as the size of the frame and B as the number of bits in the frame, and $T_{min}$ the threshold. Define the indicator function $T_n$ as

$$T_n = \begin{cases} 1 & \text{if } |(B * F_n) - (B * F_{n-1})| > T_{min} \\ 0 & \text{Otherwise} \end{cases}$$

The indicator function of the frame corresponds to a scene change. For MPEG coded videos, the ratio of the maximum peak rate to average rate may vary significantly from one video frame to another. If static allocation mechanism is used to cache MPEG video frames, it is hard to guarantee QoS while keeping buffer utilization high. For instance, cache utilization will be low if size of the cache allocated is equal to the peak rate, while the delay or the data loss rate (DLR) will be high if the average number of cache blocks is allocated. It is known from the Fig.3 that the frame size does not vary much during a scene, and hence, the following SC-caching algorithm is proposed:

**Algorithm : SC-Caching**

1. Identify the scene change
   If $\{|(B * F_n) - (B * F_{n-1})| > T_{min}\}$
     Scene change has occurred
2. If a scene change occurs, determine and allocate cache blocks for I ($CB_I$), B ($CB_B$) and P ($CB_P$) frames.
   For i = 1 to N
     $CB_I = [B*F_I]$, $CB_B = [B*F_B]$, $CB_P = [B*F_P]$.
3. Initiate a process of negotiation to acquire the required number of cache blocks for I, B and P frames.
4. Go to Step 1.

While downloading the video stream from *MMS*, $d_1$ and $d_2$ must be calculated according to which *pref-1* is replicated and *pref-2* is cached at L PSs and TR respectively. Number of cache blocks to be allocated for *I, B* and *P* frames should also be calculated by CA and assigned. This approach improves the cache utilization with no data loss. So the exact size of each frame (*I,B* or *P*) can be determined and allocated which achieves the complete cache utilization, but this may increase the allocation and reallocation overhead.

## 5 SIMULATION MODEL

In our simulation model, The VoD cluster-LPSG consists of a single *MMS* and a group of 6 *TRs*. Each of these *TR* is in turn connected to a set of 6 *PSs*. Each PS is connected to a cloud of 50 clients. Each PS has 1.5 Gbs outgoing network bandwidth. The storage capacity of each PS is 60 GB to 200 GB. The LPSG contained 300 videos with duration (size) ranged from 20 minutes to 120 minutes. The aggregate storage capacity of LPSG ranged from 380 to 600 replicas. The requests arrivals were generated by a poison process with arrival rate λ. The request distribution and video popularity distributions follows zipf-like distribution with the skew parameter ө=0.75. The user request rate at each *PS* is 35-50 requests per minutes. The ratio of cache sizes at different elements like *MMS, TR* and *PS* is set to $C_{MMS} : C_{TR} : C_{PS} = 10:4:2$ and transmission delay between the proxy and the client, proxy to proxy and *TR* to *PS* as 100ms, transmission delay between the main server and the proxy as 1200ms, transmission delay between tracker to tracker 300ms, the size of the cached *[(pref-1)+(pref-2)]* video as 280MB to 1120MB(25min – 1hr) in proportion to its popularity. Our simulation employed a simple admission control, in which a request is rejected if the requested video is currently available neither at LPSG nor at NBR[LPSG]. Also in case the video could not be accommodated even after downloading it from MMS. We use the client rejection rate ($N_{rej}$), the video miss rate (*VMR*) and the average initial access latency ($I^{a}d$) as parameters to measure the performance of our proposed approach more correctly.

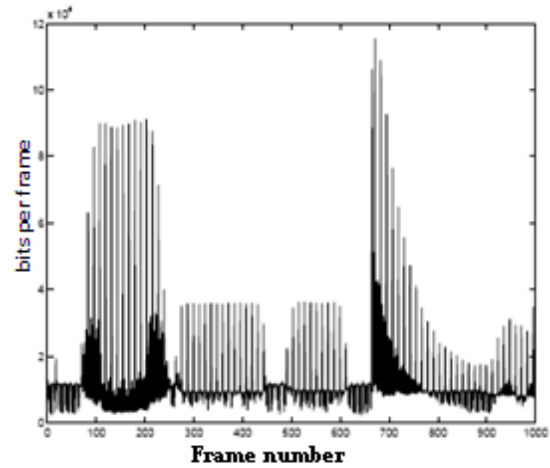

Fig 3. video traffic Sample

## 5 PERFORMANCE EVALUATION

Simulation model has been evaluated several times. The results presented below are an average of several simulations conducted. The result shows an improved performance over the existing classical replication with round robin (RR) placement and Zipf replication with smallest load first (SLFA) algorithms.

**Client rejection ratio:** Fig.4 shows the client rejection rate of our proposed system, which is very less when compare to the classical replication with RR placement (CR-RR) and Zipf replication with SLFA (ZipfR-SLFA). We can observe that, proposed regional popularity based replication and placement (RPR-P) algorithm has achieved 13% reduction in client rejection rate over zipf replication and nearly 34% when compare to classical replication. This is achieved by replicating the *pref-1* of more number of videos based on their local popularity. Even in case of cache miss, sharing of the video replicas among the PS of LPSG makes the system to achieve this significant reduction in client rejection rate.

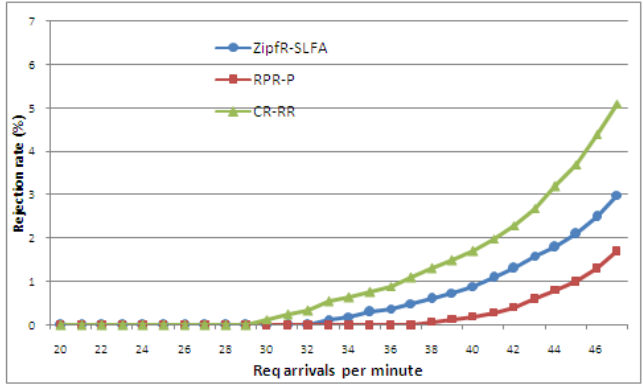

Fig. 4. Avg Rejection rate of RPR-P, ZipfR-SLFA and CR-RR Algorithms

**Initial access delay & Cache miss rate :** Regional popularity based size $d_1$ of the replica allows all the clients who arrives within $0 - d_1$ interval to get the service immediately from PS. This is because as the request arrival rate for the video $v_i$ increases $d_1$ of $v_i$ also increases. In comparison with zipf replication and classical replication this approach reduces 25%-35% video miss rate as shown in fig.6 and it in turn has reduced 3 to 8 seconds of the average initial access delay for the users when compare to zipf replication and classical replication as shown in the fig.5. Even in rare case of cache miss at the PS, at which request arrives, sharing of the video replicas present among other PSs of the LPSG and NBR-LPSG increases the service availability.

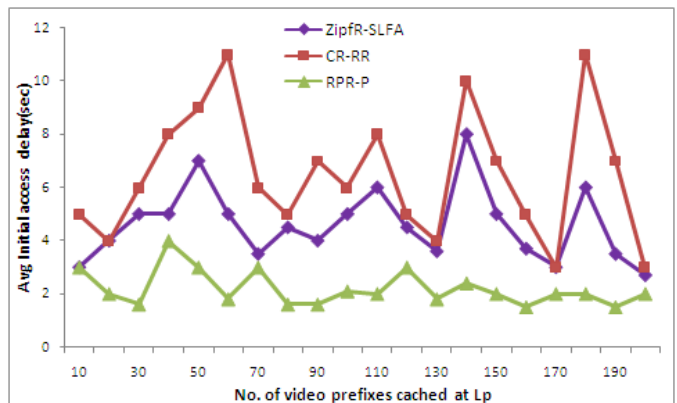

Fig. 5. Avg $I^2d$ with RPR-P, ZipfR-SLFA and CR-RR Algorithms

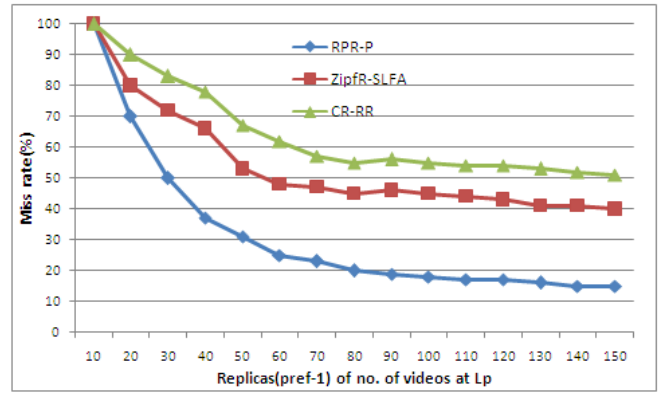

Fig. 6. Avg cache miss rate of RPR-P, ZipfR-SLFA and CR-RR Algorithms

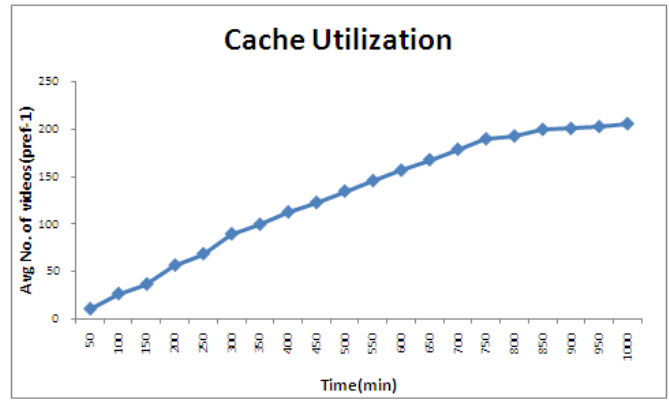

Fig. 7. Avg cache utilization in RPR-P Algorithm

**Cache utilization:** Fig.7 shows the average cache utilization of the proxy server. As we can observe from the figure, as the number of requests increases over the time, number of prefixes also increased. This enables the system to achieve high service rate.

## 5 CONCLUSION

Initial access delay, request-rejection rate, request-service rate, video miss ratio and the system throughput are the most important parameters in multimedia communication networks. By emergence of new multimedia services, the need for proxy caching algorithm is essential to guarantee QoS in interactive video delivery system. We have proposed the local popularity based video replication and placement algorithm for distributed VoD architecture. The simulation results showed that the proposed algorithm has significantly increased the request service rate, reduced the request-rejection rate and video miss rate. In addition the SC-caching algorithm allowing the system to cache more replicas of most popular video prefixes based on their local popularity at respective PSs with load sharing algorithm has improved the system throughput efficiently.

**M Dakshayini.** holds M.E in computer science. She has one and a half decades of experience in teaching field. She has published many papers. Currently she is working as a teaching faculty in the department of Information science and engineering at BMS College Of Engineering , Bangalore ,India.

**Dr.T R Gopalakrishnan Nair** holds M.Tech. (IISc, Bangalore) and Ph.D. degree in Computer Science. He has 3 decades experience in Computer Science and Engineering through research, industry and education. He has published several papers and holds patents in multi domains. He won the PARAM Award for technology innovation. Currently he is the Director of Research and Industry in Dayananda Sagar Institutions, Bangalore, India.